\documentclass[twocolumn]{aastex631}

\usepackage{float}
\usepackage{amsmath} % or simply amstext

\newcommand\au{\mathrm{au}}
\newcommand\pc{\mathrm{pc}}

\newcommand\Msun{M_{\odot}}

\newcommand{\Angstrom}{\textup{\AA}}

\defcitealias{Edenhofer+2023}{E23}
\defcitealias{Leike+2020}{L20}

\received{\today}
\revised{\today}
\accepted{\today}

\graphicspath{{./}{plots/}}

\begin{document}

%\title{Optical and near-infrared \textit{Hubble} Space Telescope observations of the edge-on protoplanetary disk IRAS 23077+6707 (``Dracula's Chivito")}
\title{\textit{Hubble} reveals complex multi-scale structure in the edge-on protoplanetary disk IRAS~23077+6707}

\author[0000-0002-5688-6790]{Kristina Monsch}
\affiliation{Center for Astrophysics $\vert$ Harvard \& Smithsonian, 60 Garden Street, Cambridge, MA 02138, USA}

\author[0000-0002-4248-5443]{Joshua Bennett Lovell}
\affiliation{Center for Astrophysics $\vert$ Harvard \& Smithsonian, 60 Garden Street, Cambridge, MA 02138, USA}

\author[0000-0002-2805-7338]{Karl R. Stapelfeldt}
\affiliation{Jet Propulsion Laboratory, California Institute of Technology, Mail Stop 321-100, 4800 Oak Grove Drive, Pasadena, CA 91109, USA}

\author[0000-0003-2253-2270]{Sean M. Andrews}
\affiliation{Center for Astrophysics $\vert$ Harvard \& Smithsonian, 60 Garden Street, Cambridge, MA 02138, USA}

\author[0009-0004-1845-3398]{Ammar Bayyari}
\affiliation{Department of Physics and Astronomy, University of Hawaii, Honolulu, HI 96822, USA}

\author[0000-0003-2014-2121]{Alice S. Booth}
\affiliation{Center for Astrophysics $\vert$ Harvard \& Smithsonian, 60 Garden Street, Cambridge, MA 02138, USA}

\author[0000-0002-9540-853X]{Adolfo S. Carvalho}
\affiliation{Department of Astronomy, MC 249-17, California Institute of Technology, Pasadena, CA 91125; USA}

\author[0000-0002-1783-8817]{John H. Debes}
\affiliation{AURA for ESA, Space Telescope Science Institute, 3700 San Martin Dr., Baltimore, MD 21218, USA}

\author[0000-0002-0210-2276]{Jeremy J. Drake}
\affiliation{Lockheed Martin, 3251 Hanover St, Palo Alto, CA 94304, USA}

\author[0009-0008-3966-0678]{Joshua W. J. Earley}
\affiliation{University of Southampton, University Rd, Southampton, UK}

\author[0000-0002-8791-6286]{Cecilia Garraffo}
\affiliation{Center for Astrophysics $\vert$ Harvard \& Smithsonian, 60 Garden Street, Cambridge, MA 02138, USA}

\author[0000-0002-3490-146X]{Garrett K. Keating}
\affiliation{Center for Astrophysics $\vert$ Harvard \& Smithsonian, 60 Garden Street, Cambridge, MA 02138, USA}

\author[0000-0003-1799-1755]{Michael L. Sitko}
\affiliation{Department of Physics, University of Cincinnati, Cincinnati, OH 45221, USA}
\affiliation{Center for Extrasolar Planetary Systems, Space Science Institute, 4750 Walnut Street, Suite 205, Boulder, CO 80301, USA}

\author[0000-0003-1526-7587]{David J. Wilner}
\affiliation{Center for Astrophysics $\vert$ Harvard \& Smithsonian, 60 Garden Street, Cambridge, MA 02138, USA}

\begin{abstract}

We present high-resolution ($\lesssim 0\farcs1$) \textit{Hubble} Space Telescope (HST)/Wide Field Camera 3 (WFC3) imaging of the near edge-on ($i{\sim}80^\circ$) protoplanetary disk IRAS~23077+6707 (``Dracula’s Chivito") obtained across six broadband filters spanning $0.4$--$1.6\,\micron$.
These observations unveil the scattered light from this unusually large disk (${\sim}14''$, or ${\sim}$4200\,au at 300\,pc) in remarkable detail, revealing a rich tapestry of substructures, including brightness asymmetries and signatures of dynamical activity.
Extended filaments are detected extending ${\sim}10''$ from the northern edges of both nebulae, while no comparable southern features are observed.
%A strong east–west brightness contrast becomes increasingly prominent toward shorter wavelengths, consistent with wavelength-dependent scattering by small dust grains.
In addition to large-scale asymmetries, the disk exhibits prominent wispy features that extend well above the midplane and are visible in all filters, suggesting a complex, possibly turbulent outer disk atmosphere shaped by infall, dynamical stirring, or gravitational instability.
%\red{At $1.6\,\micron$, a compact source appears at the dynamical center derived from NOEMA C$^{18}$O kinematics, which could trace the central star.} 
The central dark lane narrows from optical to near-IR wavelengths, and high-resolution millimeter data reveal compact midplane emission. Although our radiative transfer simulations show that the current data cannot yet distinguish between dust settling and no-settling scenarios, they underscore the need for resolved mid-infrared observations of this unique system. IRAS~23077+6707 thus represents a rare and valuable laboratory for studying the vertical structure, asymmetries, and evolutionary state of protoplanetary disks.

\end{abstract}

\section{Introduction} 
\label{sec:introduction}

Protoplanetary disks, a natural by-product of star formation, serve as both the birthplace and material reservoir for planet formation \citep[see reviews by][]{WilliamsCieza2011, Andrews+2020}. Gaining insight into their internal structure is key to understanding the mechanisms that govern the formation of planetary systems.

Among these disks, those viewed at very high inclinations ($i \gtrsim 80^\circ$)---commonly referred to as edge-on disks (EODs)---provide a particularly valuable perspective. In such systems, the disk's dust and gas act as an effective coronagraph, blocking direct light from the central star and revealing the disk in silhouette. This results in a distinctive appearance in optical and infrared (IR) images: a central dark lane flanked by bright scattered light. The geometry of EODs enables direct observations of the radial and vertical structure of disks, providing critical insights into dust growth and settling \citep{Wolf+2003, Villenave+2020, Duchene+2024, Tazaki+2025}, molecular stratification \citep[][]{RuzRodriguez+2021, Sturm+2023, Arulanantham+2024, Dartois+2025, Pascucci+2025}, and temperature gradients \citep{vant'Hoff+2018, Guilloteau+2025} as a function of both scale height and radius.
The availability and size distribution of dust, its interaction with gas, and its vertical structure all play a vital role in shaping the planetary systems that emerge from disks \citep{Bae+2023_PP7}.

The first spatially resolved images of EODs were obtained with the \textit{Hubble} Space Telescope (HST), which pioneered optical scattered light imaging of these objects at sub-arcsecond resolution.
Since most protoplanetary disks span only a few arcseconds on the sky, resolving these structures at visible wavelengths requires extremely high angular resolution, as demonstrated by iconic EODs such as HH~30 \citep{Burrows+1996, Stapelfeldt+1997, WatsonStapelfeldt2004}, IRAS~04302+2247 \citep[``Butterfly Star'',][]{LucasRoche1997, Padgett+1999}, 2MASS J04202144+2813491 \citep[Tau042021,][]{Duchene+2014, Stapelfeldt+2014} or IRAS~18059-3211 \citep[``Gomez's Hamburger'',][]{Ruiz+1987, Bujarrabal+2008, Bujarrabal+2009}.

Among all known EODs, IRAS~23077+6707 (``Dracula’s Chivito", hereafter IRAS23077) stands out: it is a recently identified Herbig Ae/Be system with an exceptionally large angular extent of ${\sim}14''$ and striking asymmetries.
First recognized as a protoplanetary disk in 2024 through combined Pan-STARRS (PS1) and Submillimeter Array (SMA) observations \citep{Berghea+2024, Monsch+2024}, IRAS23077 represents a rare and distinctive addition to the population of EODs. 
It exhibits pronounced brightness asymmetries in both scattered light and millimeter emission. While the strong east–west contrast can be attributed to the disk's near edge-on inclination of ${\sim}80^\circ$---with the western nebula tilted toward the observer and thus appearing brighter---the eastern lobe also shows a distinct north–south asymmetry, with the southern side brighter by a factor of three  \citep{Monsch+2024}.
High-resolution SMA and Northern Extended Millimetre Array (NOEMA) observations presented by \citet{Lovell+2025} further reveal an opposing 50\% enhancement in millimeter emission compared to the scattered light, which they showed was consistent with IRAS23077 hosting a moderately eccentric ($e\approx0.26$) disk.
These multi-wavelength asymmetries in IRAS23077 therefore highlight the need for high-resolution observations across the electromagnetic spectrum to fully characterize this unique edge-on system. 

In this Paper, we present the first sub-arcsecond optical and near-IR imaging of IRAS23077, obtained with HST/WFC3 from 0.4–$1.6\,\micron$. These data offer an unprecedented view of the system's scattered light structure and provide a crucial step towards understanding its complex morphology. 
The retrieval and reduction of the observational data are described in \S\ref{sec:observations}. We present and discuss our results in \S\ref{sec:results} and provide our conclusions in \S\ref{sec:conclusions}. 

\section{Observations} 
\label{sec:observations}

\begin{figure*}[t!]
    \centering
    \includegraphics[width=1.\linewidth, clip,trim={0.65cm 0.55cm 0.0cm 0.0cm}]{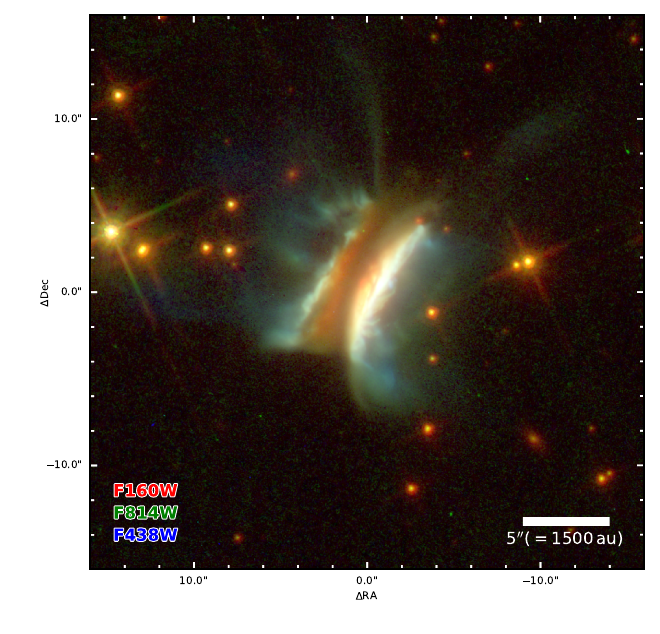} 
    \caption{Three-color composite of IRAS23077 using the HST/WFC3 filters F160W (red), F814W (green), and F438W (blue). 
    A $5''$ scale bar is shown in the lower right, which at a distance of $300\,\pc$ (see \S\ref{sec:results}) corresponds to $1500\,\au$. North is up, east is left.}
    \label{fig:RGB}
\end{figure*}

We observed IRAS23077 as part of HST-GO-17751 (PI: K. Monsch, Co-PI: J. B. Lovell) for three orbits using the UVIS and IR channels of the HST/WFC3 instrument \citep[][]{MacKenty+2010, MacKenty+2014, HST_WFC3_Instrument_Handbook}. 
The observations were carried out on February 08, 2025 and used the following six broadband filters: F438W ($0.4\,\micron$), F606W ($0.6\,\micron$), F814W ($0.8\,\micron$), F105W ($1.05\,\micron$), F125W ($1.25\,\micron$) and F160W ($1.60\,\micron$). This set of filters was selected to cover the full breadth of the optical spectrum previously studied by PS1, as well as to obtain the first spatially resolved near-IR images of IRAS23077. The observations are summarized in Table~\ref{tab:obs_summary_phot}. 

\begin{deluxetable*}{ccccccccc}[t]
\tablecaption{Summary and photometry of the newly obtained HST and archival millimeter (SMA/NOEMA) observations of IRAS23077.\label{tab:obs_summary_phot}}
\tablecolumns{9}
\tablehead{
\colhead{Filter} & \colhead{Instrument} & \colhead{$t_\mathrm{int}$} & \colhead{Obs. IDs} & \colhead{$F_{\nu}$} & \colhead{$m_{\mathrm{AB}}$} & \colhead{Flux ratio} & \colhead{$d_\mathrm{disk}$} & \colhead{Reference} \\
\colhead{} & \colhead{} & \colhead{} & \colhead{} & \colhead{(mJy)} & \colhead{(mag)} & \colhead{} & \colhead{($''$)} & \colhead{} }
\startdata
F438W & WFC3/UVIS & 2100\,s & ifhn01veq, ifhn01010 & $5.1 \pm 0.1$ & $14.6 \pm 0.3$ & $14.3 \pm 0.9$ & $3.1 \pm 0.1$ & \textit{this work} \\
F606W & WFC3/UVIS & 600\,s  & ifhn01020            & $7.4 \pm 0.2$ & $14.2 \pm 0.3$ & $10.1 \pm 0.6$ & $3.1 \pm 0.1$ & \textit{this work}\\
F814W & WFC3/UVIS & 1020\,s & ifhn01030, ifhn01vmq & $9.3 \pm 0.2$ & $14.0 \pm 0.3$ & $7.8 \pm 0.4$ & $2.9 \pm 0.1$ & \textit{this work}\\
F105W & WFC3/IR   & 648\,s  & ifhn01050            & $11.9 \pm 0.2$ & $13.7 \pm 0.3$ & $6.5 \pm 0.4$ & $2.6 \pm 0.1$ & \textit{this work}\\
F125W & WFC3/IR   & 648\,s  & ifhn01040            & $14.1 \pm 0.3$ & $13.5 \pm 0.3$ & $7.3 \pm 0.5$ & $2.3 \pm 0.1$ & \textit{this work}\\
F160W & WFC3/IR   & 648\,s  & ifhn01060            & $19.5 \pm 0.4$ & $13.2 \pm 0.3$ & $8.1 \pm 0.5$ & $2.1 \pm 0.2$ & \textit{this work}\\
1.3\,mm & SMA   & 18.3\,h  & --            & $49.3\pm5.5$ & $12.2\pm0.1$ & -- & $1.63 \pm 0.02$ & \citet{Lovell+2025}\\
2.7\,mm & NOEMA/USB   & 17.6\,h  & --            & $3.17\pm0.34$ & $15.2\pm0.1$ & -- & $1.07 \pm 0.02$ & \citet{Lovell+2025}\\
3.1\,mm & NOEMA/LSB   & 17.6\,h  & --            & $2.01\pm0.22$ & $15.6 \pm0.1$ & -- & $1.21 \pm 0.02$ & \citet{Lovell+2025}\\
\enddata
%\vspace{-0.6cm}
\tablecomments{All HST observations were obtained on 2025 February 08 and were centered on $\mathrm{RA (ICRS)} = 23^\mathrm{h}09^\mathrm{m}43.645^\mathrm{s}$ and $\mathrm{Dec (ICRS)} = +67^\circ23'38\farcs94$. Flux densities and AB magnitudes in the HST images were measured in a $r=7''$ aperture centered on IRAS23077's sky coordinates (see \S\ref{sec:results_photometry}). Peak-to-peak brightness ratios and dust lane thicknesses were measured along cuts perpendicular to the disk midplane (see \S\ref{sec:results_asymmetries} and \S\ref{sec:results_midplane} for details). We additionally report millimeter photometry and midplane thicknesses measured from SMA/NOEMA data.}
%\vspace{-0.8cm}
\end{deluxetable*}

After the observations were performed, we downloaded the raw data products from the \textit{Barbara A. Mikulski} Archive for Space Telescopes (MAST\footnote{\url{https://archive.stsci.edu/}}) at the Space Telescope Science Institute. 
The specific observations analyzed here can be accessed via \dataset[doi: 10.17909/mt6j-a854]{https://doi.org/10.17909/mt6j-a854}. 
We recalibrated these observations using the HST calibration pipeline \texttt{CALWF3} \citep[v3.7.2, see][for details]{HST_Data_Handbook_v6} using the default settings and the latest set of CRDS reference files (v12.0.7) as of April 1, 2025. The recalibrated data products can be downloaded from Zenodo\footnote{\url{https://doi.org/10.5281/zenodo.15530666}}. 
For the IR-images, we used the \texttt{\_drz.fits} data products for the analysis, which are calibrated exposures corrected for geometric distortion. 
For the UVIS-images, we used the \texttt{\_drc.fits} files, which are additionally corrected for Charge Transfer Efficiency (CTE) loss.\footnote{See Table~1 in \url{https://hst-docs.stsci.edu/wfc3dhb/chapter-2-wfc3-data-structure/2-1-types-of-wfc3-files} for a summary of the different HST pipeline data products.} 
HST WFC3 data products are provided in units of count rates (e$^-$/s), but can be converted into flux density units using the \texttt{PHOTFLAM} (inverse sensitivity in units of $\mathrm{erg/cm^2/\Angstrom/e^-}$) and \texttt{PHOTPLAM} (Pivot wavelength in $\Angstrom$) fits-header keywords, and in AB magnitudes via the relation: $m_\mathrm{AB} = -2.5\log F_\nu - 48.6$, where $F_\nu$ is expressed in units of $\mathrm{erg\,cm^{–2}\,s^{–1}\,Hz^{–1}}$.

\section{Results and discussion} 
\label{sec:results}

Figure~\ref{fig:RGB} presents a three-color composite image of IRAS23077, combining data from filters F438W, F606W (WFC3/UVIS) and F160W (WFC/IR). The system displays the hallmark structure of a highly inclined disk: two extended bright lobes separated by a prominent dark lane tracing the disk midplane. The disk appears significantly flared and vertically extended, with faint scattered light extending well above the disk's main body. 
A $5''$ scale bar is shown, which at a distance of $300\,\pc$ \citep[adopted by][as possible distance to IRAS23077]{Berghea+2024} corresponds to a physical extent of $1500\,\au$. We emphasize, however, that the distance to IRAS23077 is currently \textit{not} known. Due to its extended size on the sky, no accurate \textit{Gaia} parallax measurements are available. However, its northern sky position suggests that it is likely part of the Cepheus star-forming region, which has an inferred distance range of 150–$370\,\pc$ \citep{Szilagyi+2021}, and is thus consistent with the value of $300\,\pc$ adopted by \citet{Berghea+2024} based on \textit{Gaia} extinctions. 

\begin{figure*}
    \centering
    \includegraphics[width=\linewidth]{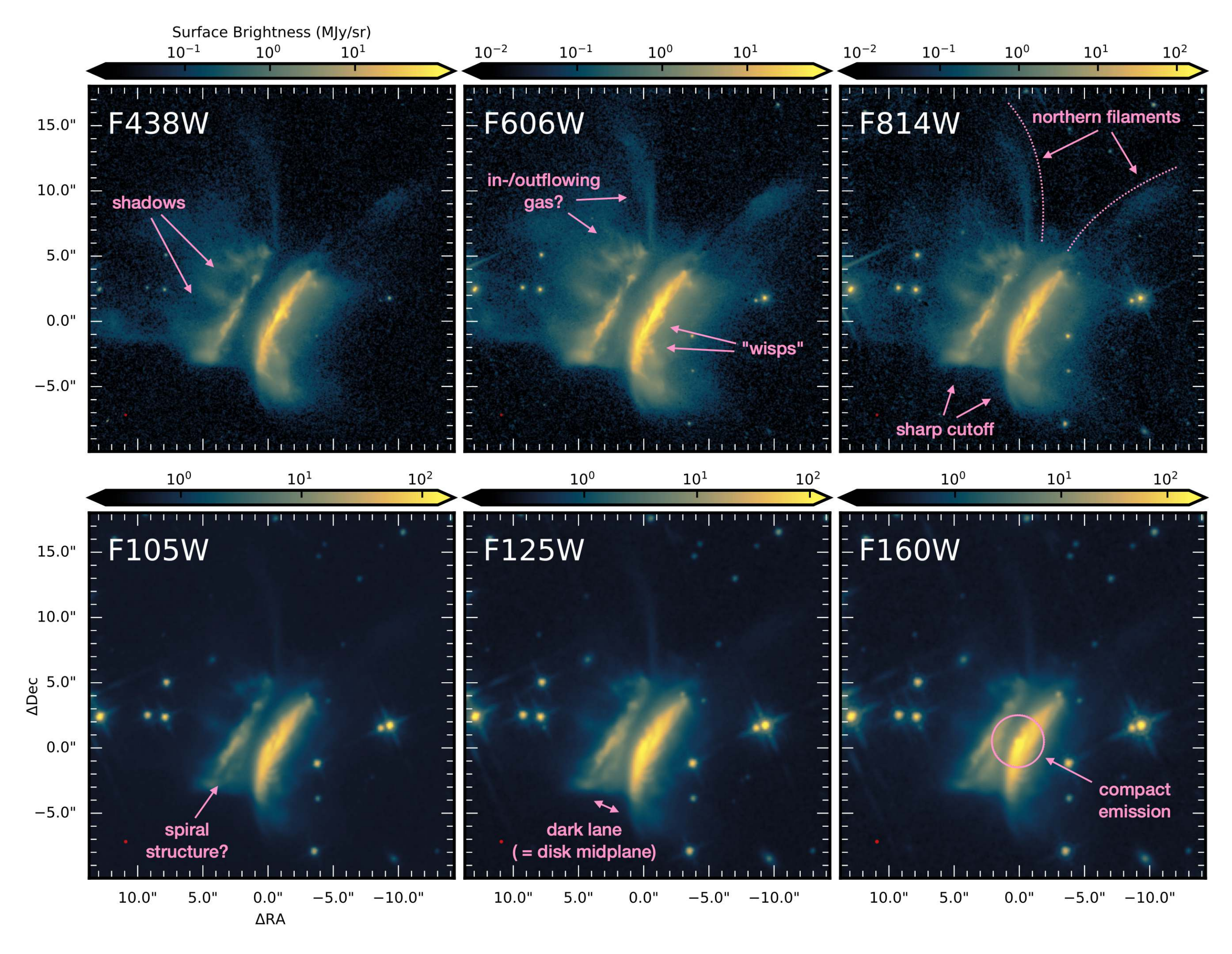}
    \caption{Image gallery of all HST observations of IRAS23077. The top row shows images taken with the WFC3/UVIS filters F438W, F606W, and F814W, while the bottom row shows images taken with the WFC3/IR filters F105W, F125W, and F160W. The respective PSF FWHMs are shown as red circles in the lower left corners. All images are centered slightly offset from IRAS23077's nominal sky position to capture the full extent of the disk and the two northern filaments. North is up and east is to the left in all panels. Major features identified in the HST images are annotated and are discussed in more detail in \S\ref{sec:results}.}
    \label{fig:HST_gallery}
\end{figure*}

Several additional morphological features are evident, including wispy structures emerging from the upper disk layers, shadowed regions, and pronounced brightness asymmetries. Further, the dark lane narrows at longer wavelengths, giving the midplane a redder appearance in the RGB composite.
To visualize how these features evolve across wavelengths, Figure~\ref{fig:HST_gallery} shows a gallery of all six HST filters used in this program. While the overall disk morphology remains consistent from optical to near-IR wavelengths, the color information reveal subtle changes in vertical extent, scattering efficiency, and internal structure. 
The following sections explore these structural characteristics in more detail.

\subsection{Vertical Extent}
\label{sec:results_verticalstructure}

IRAS23077's faint scattered light, most prominent in the optical filters, extends up to ${\sim}5''$ above the disk midplane when measured at $0.8\,\micron$ (the F814W image offers the best combination of resolution and sensitivity among all HST images in this program). 
The vertical extent was determined by measuring the distance from the 100$\,\sigma_\mathrm{rms}$ contour at the top of the dark lane to the outer 3$\,\sigma_\mathrm{rms}$ contour of the faint scattered light, averaged over four equidistant measurements.
At the distance range of the Cepheus star-forming region \citep[150–370\,pc;][]{Szilagyi+2021}, this corresponds to extreme vertical heights of $750$--$1850\,\au$.
Similarly extended, bipolar nebulosities had been previously imaged in other edge-on systems such as IRAS~04302+2247, DG~Tau~B, CoKu~Tau/1 (HST/NICMOS; \citealt{Padgett+1999, Villenave+2024a_IRAS04302+2247}) or in HOPS~136 (HST/NICMOS/WFC3; \citealt{Fischer+2014}).

IRAS23077 further exhibits prominent wispy structures that protrude well above the disk midplane, clearly visible across all six HST filters. These features include both bright material seen above the nebula and darker filaments in the foreground.
In edge-on systems, the outer disk is projected in front of the inner regions, suggesting that these extended features arise from the disk's outer radius.
Their vertical extent and irregular appearance suggest a significant level of turbulence, possibly reflecting material that is still settling or infalling. The presence of such features complicates  efforts to infer a stable disk scale height from the optical images, as a well-defined vertical pressure gradient may not yet have formed. However, without detailed modeling, it remains unclear whether the disk is dynamically young \citep[e.g.][]{Flock+2020}, stirred up by a planet/companion \citep[e.g.][]{Facchini+2018} or shaped by other processes, such as late infall \citep{Dullemond+2019, Kuffmeier+2020, Ginski+2021}, or gravitational instability \citep{LodatoRice2004, LodatoRice2005, Dong+2015, Pohl+2015, Speedie+2024}.
Nevertheless, the ubiquity and coherence of the wisps across wavelengths hint at ongoing dynamical activity in IRAS23077.

\subsection{Scattered light asymmetries}
\label{sec:results_asymmetries}

\begin{figure}[t!]
    \centering
    \includegraphics[width=0.95\linewidth]{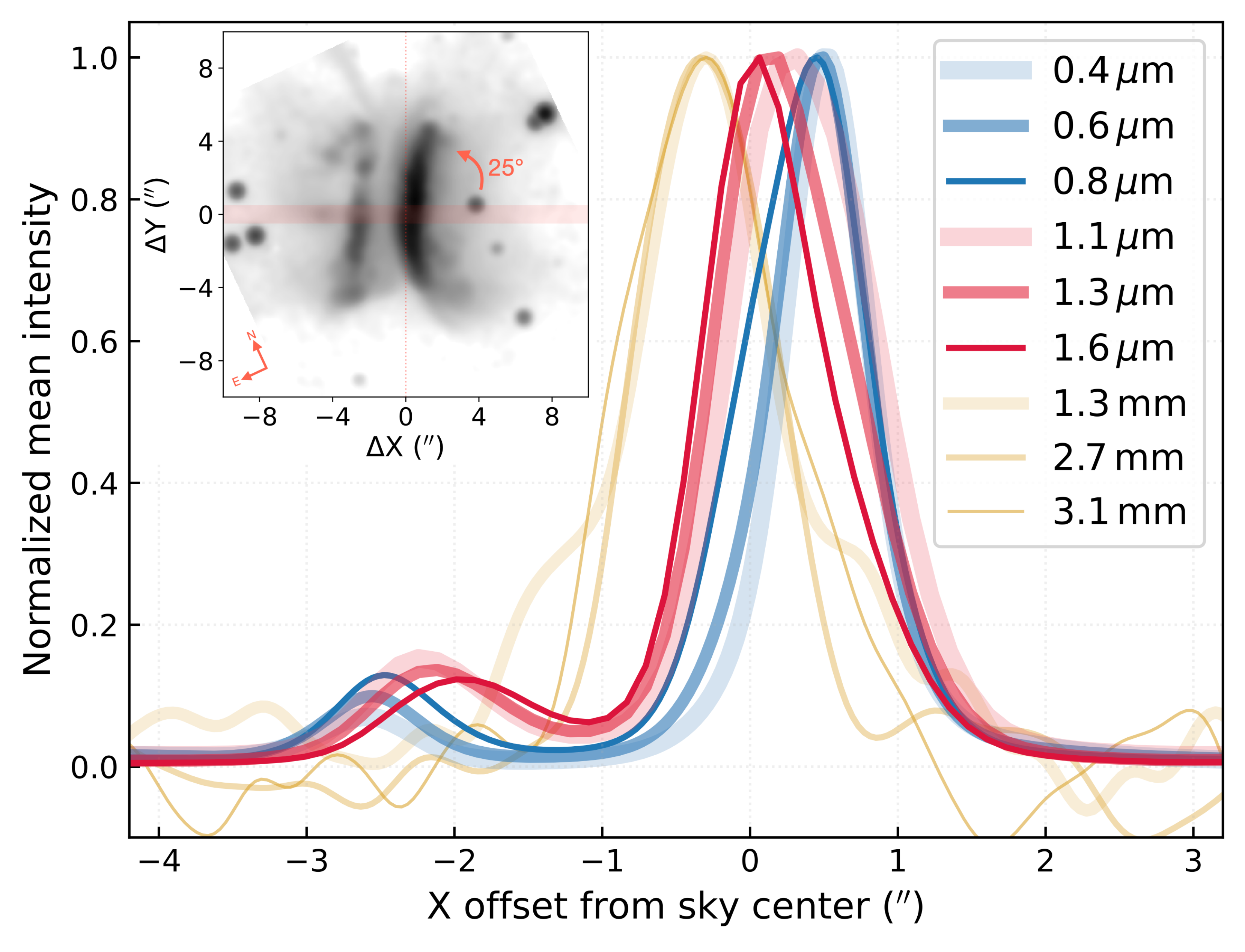}
    \caption{Vertical intensity profiles of IRAS23077. The HST images were rotated by $25^\circ$ counterclockwise to align the disk midplane vertically, and the mean emission was computed from all pixel lanes within a $1''$-wide band (red band in the cutout image in the upper left corner) perpendicular to the midplane (see \S\ref{sec:results_asymmetries} for details). Each profile is normalized to its maximum and centered on IRAS23077's sky coordinates. Blue lines show optical filters (0.4--$0.8\,\micron$), red lines the IR filters (1.1--$1.6\,\micron$), and yellow lines the SMA/NOEMA millimeter profiles (1.3--$3.1\,\mathrm{mm}$; \citealt{Lovell+2025}). The bright peak at zero offset corresponds to the western nebula, the fainter peak to the eastern nebula, and the gap between them marks the central dark lane. }
    \label{fig:vertical_profiles}
\end{figure}

The newly revealed features in IRAS23077 include shadowed regions
%and possible in-/outflowing material 
in the eastern lobe, as well as the previously mentioned wispy structures emerging from the disk's outer radius. At IR wavelengths, an additional structure within the eastern lobe becomes visible (see, for instance, the F105W image in Figure~\ref{fig:HST_gallery}). This is likely due to the lower opacity in the IR, which allows deeper layers closer to the disk midplane to be probed. In these data, the eastern lobe shows both the outer region seen in the UVIS images and an inner component that is only revealed at longer wavelengths.
Similar features have been observed in other EODs, such as HH30 \citep{Burrows+1996, Cotera+2001, WatsonStapelfeldt2007}, and have been proposed to trace spiral structures \citep[][]{Norfolk+2022, Tazaki+2025}. Alternatively, it could correspond to a shadowed region, possibly cast by a misaligned inner disk onto the outer disk, as suggested in other systems \citep[e.g. SU Aur, cf.][and see also \citealt{Ansdell+2020}]{Labdon+2019, Ginski+2021}.

We repeated the measurement of the intensity ratios between the eastern and western nebulae from \citet{Monsch+2024}. For this analysis, we rotated each HST image by $25^\circ$ counterclockwise to align the disk midplane with the vertical axis. This corresponds to a disk position angle of $\mathrm{PA}=335^\circ$, as recently measured by \citet{Lovell+2025} in high-resolution millimeter observations of IRAS23077. 
However, owing to the asymmetric optical/IR morphology of IRAS23077, this rotation does not produce a perfectly vertical alignment, and the two bright nebulae remain slightly tilted with respect to one another. We tested several plausible position angles, but none yielded a fully satisfactory alignment. This lateral asymmetry was already noted by \citet{Berghea+2024} based on the lower-resolution PS1 imaging, potentially pointing to a tilted inner disk region. Indeed, \citet{Villenave+2024a_IRAS04302+2247} found that many EODs observed in scattered light exhibit comparable offsets, supporting this interpretation also for IRAS23077.

All HST images were then smoothed with a symmetric 2D Gaussian kernel with a FWHM of $0\farcs5$ in order to suppress small-scale substructures (e.g., the wisps) that would otherwise introduce multiple peaks in the vertical profiles. We checked that this smoothing does not affect our key measurements: while the profiles become somewhat broader and peak intensities change, the peak-to-peak separation and flux ratios remain essentially unchanged. We then extracted the profiles by averaging pixel lanes within a $1''$-wide band centered on the disk midplane (see the red band in the cutout image in  Figure~\ref{fig:vertical_profiles}). This approach maximizes the signal-to-noise ratio and reduces the influence of disk flaring and shadowing on the profiles.

The resulting profiles are shown in Figure~\ref{fig:vertical_profiles}.
The bright peak near zero offset corresponds to the western nebula, while the fainter peak traces the eastern nebula; the gap between them marks the central dust lane.
Because the images are centered on IRAS23077's sky coordinates, the midplane does not fall exactly at zero offset but is slightly shifted toward negative values--consistent with the disk being slightly inclined away from the observer.
Also, a gradual narrowing of the midplane is evident toward longer wavelengths: in the near-IR filters, the central gap becomes shallower and the two peaks move closer together compared to the optical bands.
This trend continues in the SMA and NOEMA profiles, whose peaks lie very close to zero offset, as the millimeter emission directly traces the narrow disk midplane and is therefore less affected by optical depth and inclination effects.

\begin{figure}
    \centering
    \includegraphics[width=\linewidth]{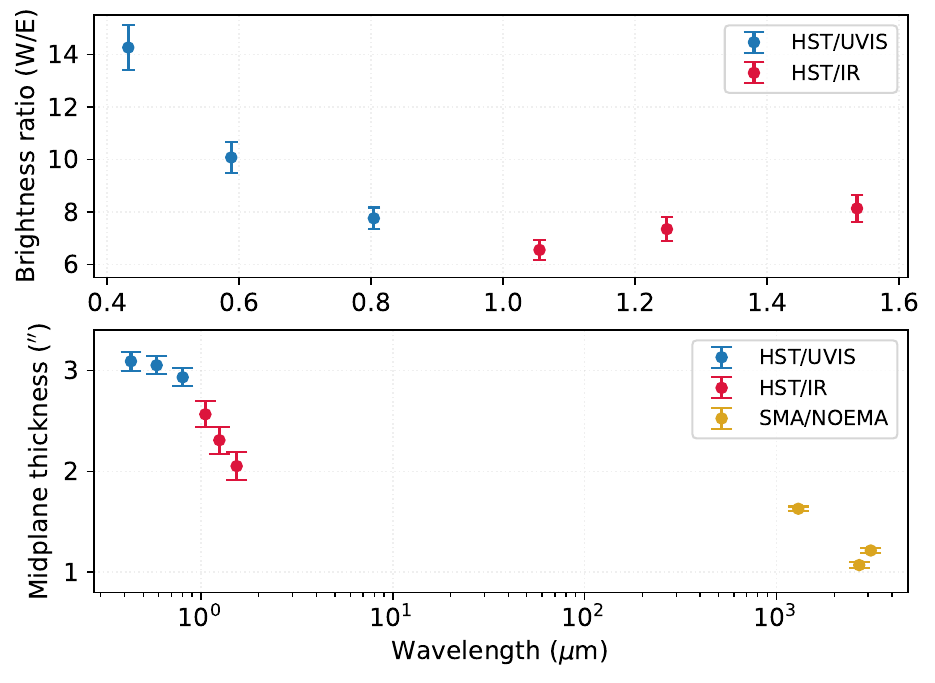}
    \caption{Peak-to-peak brightness ratio (top) and midplane thickness (bottom) measurements as a function of wavelength for IRAS23077.}
    \label{fig:obs_midplane_ratio}
\end{figure}

We find peak-to-peak flux ratios between the western and eastern lobes of $14.3\pm0.9$, $10.1\pm0.6$, $7.8\pm0.4$, $6.5\pm0.4$, $7.3\pm0.5$, and $8.1\pm0.5$ in the F438W, F606W, F814W, F105W, F125W, and F160W filters, respectively (see Figure~\ref{fig:obs_midplane_ratio}). Uncertainties are 1$\sigma$ estimates based on parametric bootstrapping with 1000 Poisson realizations of the averaged profiles, accounting for photon noise in each pixel. 
These measurements show that the brightness asymmetry increases significantly toward shorter wavelengths, but also exhibits a modest rise in the IR, possibly due to compact emission near the disk center (see \S\ref{sec:results_centralsource}).

The decreasing peak-to-peak ratios with wavelength in the HST data likely reflect a combination of stronger Rayleigh scattering by sub-micron grains in the disk's upper layers and geometric effects from the system's inclination and scattering phase function, which preferentially brighten the near-side nebula.
In the optical, additional attenuation of far-side scattered light in the optically thick disk further enhances the contrast, whereas at longer wavelengths reduced scattering efficiency and lower optical depth diminish the asymmetry.

\subsection{Midplane Thickness}
\label{sec:results_midplane}

The HST observations reveal a wavelength-dependent vertical structure in the disk, which manifests in EODs as a decreasing vertical extent and a narrowing of the dark lane at longer wavelengths.
The effect can be quantified using the profiles shown in Figure~\ref{fig:vertical_profiles}, which allow the measurement of the ``midplane thickness'' as a function of wavelength. Comparing these measurements with theoretical disk models provides constraints on grain opacities, size distributions, and the degree of vertical dust settling, offering direct insight into the extent of grain growth \citep[e.g.][]{D'Alessio+2001, D'Alessio+2006, WatsonStapelfeldt2004, Duchene+2024, Tazaki+2025}. 

For each HST filter, we determined the midplane thickness by identifying the two intensity peaks in the vertical emission profiles (corresponding to the eastern and western nebulae, respectively) and measuring their peak-to-peak separation. This yields thicknesses of $3\farcs1\pm0\farcs01$, $3\farcs11\pm0\farcs01$, $2\farcs9\pm0\farcs1$, $2\farcs6\pm0\farcs1$, $2\farcs3\pm0\farcs1$, and $2\farcs1\pm0\farcs2$ for filters F438W, F606W, F814W, F105W, F125W, and F160W, respectively (see the lower panel in Figure~\ref{fig:obs_midplane_ratio}). We performed analogous measurements for the SMA and NOEMA millimeter data from \citet{Lovell+2025}, where the emission is, however, only single-peaked. In these cases, Gaussian functions were fitted to the vertical profiles, and the disk thickness was then defined as the fitted FWHM, with $1\sigma$ uncertainties derived from the fits.

The observed decrease in midplane thickness with increasing wavelength indicates a wavelength-dependent vertical structure in the scattering material. This trend may reflect either a true vertical stratification of grain sizes through dust settling \citep{Dubrulle+1995, DullemondDominik2004, Watson+2007_PP5} or simply an opacity-driven effect in a well-mixed disk. In a vertically mixed disk without dust settling, lower opacities at longer wavelengths shift the $\tau{=}1$ surface closer to the midplane, naturally producing a thinner dark lane. This effect was demonstrated by \citet{WatsonStapelfeldt2004} for HH30, where no vertical settling was required to explain the narrowing of HH30's dust lane at longer wavelengths. 

\begin{figure}
    \centering
    \includegraphics[width=1\linewidth]{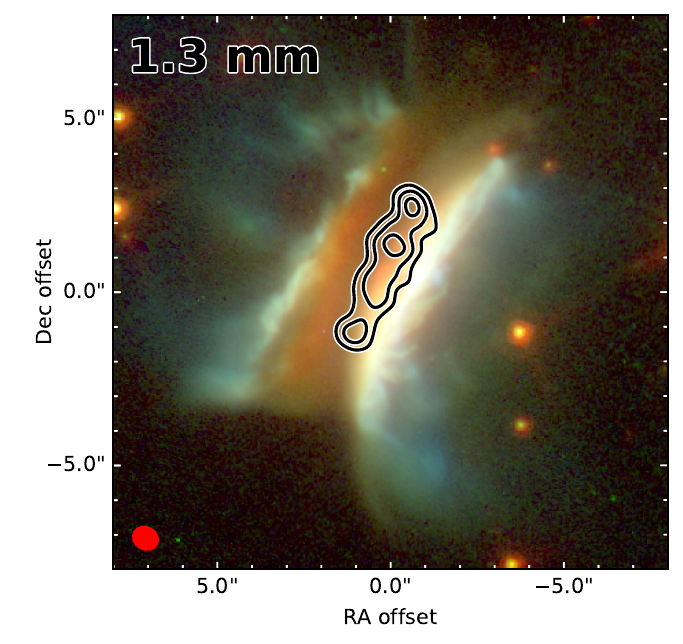}
    \caption{Comparison of optical/IR scattered light and dust continuum emission in IRAS23077. The contours show the SMA 1.3\,mm/225\,GHz emission from \citet{Lovell+2025} at $[8,12,16]\times \sigma_\mathrm{rms}$ levels. In the lower left corner, the synthesized beam of the SMA data is shown ($0\farcs80\times0\farcs69$, BPA$=61^\circ$). North is up, east is left. }
    \label{fig:HST_with_SMA}
\end{figure}

Therefore, scattered light imaging alone cannot unambiguously diagnose dust settling. Direct evidence requires tracing larger grains with high-resolution millimeter continuum data. In this regime, a fully mixed disk would appear more vertically extended and flared, whereas in a settled disk the larger dust grains are expected to be more strongly concentrated toward the disk midplane, and thus producing a ``cigar-shaped" morphology \citep[see, e.g., Fig.~13 in][]{Duchene+2024}.

The millimeter observations of IRAS23077 indeed reveal that the dust continuum emission is narrowly confined to the disk midplane (see Figure~\ref{fig:HST_with_SMA}, which shows the dust continuum overlaid on the HST image). 
At the resolution of these data ($\sim 0\farcs7$--$0\farcs8$), the millimeter emission shows no obvious extended nor flared morphology, which could suggest that vertical settling of large grains may already be underway in IRAS23077. 
However, only through direct comparison with theoretical disk models which explicitly include dust settling can this hypothesis be tested. 
We therefore performed a suite of radiative transfer simulations exploring the degree of dust settling in IRAS23077. We provide the full description of our modeling framework and a more detailed discussion of these results in Appendix~\ref{sec:appendix_modelling}, while we will only summarize the main results below.

The models show that it is currently not possible to distinguish between settled and unsettled scenarios: both produce nearly identical scattered light morphologies and radial profiles, with only subtle differences such as slightly more extended wings of the millimeter emission in the unsettled case (see Figure~\ref{fig:obs_vs_model}). These differences become most apparent at millimeter wavelengths, yet the available SMA and NOEMA observations lack the resolution required to discriminate between the modeled scenarios. We note that although IRAS23077 cannot be observed with ALMA due to its northern declination, future observations, as for instance with the next-generation Very Large Array (ngVLA) offer a promising avenue to probe settling in this disk \citep[e.g.,][]{Andrews+2018_ngVLA, Ricci+2018}.

The lack of wavelength coverage between the optical and the millimeter is an additional limitation. This intermediate regime is where emission from micron- to tens-of-micron-sized grains starts to become sensitive to vertical settling, yet no resolved data exist at these wavelengths for IRAS23077. 
Although in other protoplanetary disks the combination of sub-/millimeter and mid-IR imaging with ALMA and JWST has been essential for probing the degree of settling \citep[cf.][]{Duchene+2024}, in the case of IRAS23077, only the addition of mid-IR JWST imaging offers a viable path toward investigating dust settling in its disk with the currently available facilities.

\subsection{Northern Filaments and Sharp Cutoff in the South}
\label{sec:results_filaments}

The northern filamentary structures first identified in PS1 imaging are clearly recovered in the HST data, extending up to ${\sim}10''$ above the disk midplane. Although confined to the northern side, they exhibit a high degree of symmetry with respect to each other. 
Compared to the main disk emission, the filaments become fainter toward longer wavelengths, which may be a result of the reduced sensitivity and higher background noise in the IR images. However, it could also reflect differences in dust grain properties or illumination. 

Similar one-sided filamentary structures have been observed in other protoplanetary disks, such as SU~Aur \citep[][]{DeLeon+2015, Akiyama+2019, Ginski+2021}, where both scattered light and ALMA data reveal extended arcs and narrow dusty filaments. \citet{Ginski+2021} interpreted these features as signatures of infalling material. In IRAS23077, the northern filaments have likewise been suggested to trace remnants of an infalling envelope \citep{Berghea+2024}, although the models in that study predict roughly symmetric structures on both sides of the disk.
From the PS1 data alone, it was unclear whether the apparent absence of southern filaments might result from limited sensitivity. However, the HST images are substantially deeper\footnote{We estimate a $5\sigma$ limiting AB magnitude of $26.03\pm0.27$\,mag for our F814W image, far deeper than the $5\sigma$ point-source depth of the Pan-STARRS $3\pi$ survey, which reaches 23.1\,mag in $i$ and 22.3\,mag in $z$ \citep[which are the filters closest in wavelength to F814W, cf.][]{PS1_Chambers+2016}.}, yet still show no evidence of southern counterparts. 
Instead, the southern end exhibits a sharp cutoff in scattered light, indicating that the asymmetry is intrinsic to the system. A pure foreground extinction scenario appears unlikely, as a background star is clearly visible south of the disk at a similar projected distance to the northern filaments.
So while the northern filaments may indeed trace material from a dispersing or infalling envelope, the southern region likely harbors the same envelope material, which is, however, not illuminated in the same way and therefore remains undetected in scattered light.

\subsection{Photometry}
\label{sec:results_photometry}

We measure IRAS23077's flux via aperture photometry, using the Python package \texttt{photutils} \citep{Bradley+2025_photutils}. For this purpose, we selected a circular ($R=7''$) aperture encompassing the disk's full extent and a nearby $20''\times20''$ noise-dominated region to estimate the background contribution. 
The circular aperture also encompasses four stars near the western nebula, with one being located right at the disk's northern upper boundary\footnote{Although we subtract their fluxes from the PSF photometry, these background stars offer promising avenues to probe ices and dust species in IRAS23077's outer disk via absorption spectroscopy \citep[e.g.,][]{Sturm+2023, Potapov+2025, vanDieshoeck+2025}.}. To correct for this contamination, we first measured the empirical median PSF FWHM from field stars in each HST image and then used the resulting PSF models to fit and subtract the emission from the four intervening stars via PSF photometry.
The resulting flux densities and AB magnitudes are listed in Table~\ref{tab:obs_summary_phot}. As expected for a pre-main sequence (PMS) star with IR excess, IRAS23077's flux rises steadily with wavelength, from $5.1 \pm 0.1$\,mJy at $0.4\,\micron$ to $19.5 \pm 0.4$\,mJy at $1.6\,\micron$, consistent with previously reported PS1 photometry \citep{Berghea+2024}.

While we detect no significant optical variability, this conclusion is based on only two widely separated epochs---PS1 imaging from roughly a decade ago and our recent HST photometry---which agree within the uncertainties. The sparse optical sampling therefore cannot rule out shorter-term or low-amplitude variations. 
In contrast, NEOWISE monitoring \citep[from the NEOWISE Reactivation Mission, see][]{Mainzer+2014} provides a decade of regular cadence and reveals a mild ${\sim}0.3$\, mag mid-IR brightening and a weak ${\sim}0.1$\,mag W1-W2 color modulation (see Fig.~\ref{fig:neowise} in Appendix~\ref{sec:appendix_neowise}). Although much smaller, this IR behavior resembles UX Ori-type variability \citep{Herbst+1994, Natta+1997, Kreplin+2016}, which has been linked to misaligned inner disks casting moving shadows \citep[e.g.,][]{Marino+2015, Benisty+2018, Facchini+2018, Ansdell+2020, Pontoppidan+2020, Kuffmeier+2021}. 
Given the sparse optical coverage and the only very weak mid-IR variability, a definitive interpretation for IRAS23077's mid-IR variability remains, however, premature.

\subsection{Detection of Point-like Emission}
\label{sec:results_centralsource}

The F160W HST image reveals a compact emission feature near the center of IRAS23077's western lobe. The peak is undetected at shorter wavelengths and is spatially coincident with the disk's dynamical center at RA$=23^\mathrm{h}09^\mathrm{m}43\fs7,\;
$Dec$ = +67^\circ23'39\farcs5$ (i.e. offsets of ($\Delta$RA, $\Delta$Dec) $= (+0\farcs18, +0\farcs58)$ relative to the nominal source position) inferred from NOEMA C$^{18}$O $J{=}1{-}0$ kinematics \citep{Lovell+2025}. 

A 1D Gaussian fit to the brightness profile across the feature yields $\mathrm{FWHM}=0\farcs78\pm0\farcs05$, substantially broader than the F160W PSF in our drizzled data ($\sim0\farcs2$) and therefore does not correspond to a point source. This indicates that the emission is not direct starlight but scattered, thermally emitted photons propagating through the disk. Radiative transfer models (see Appendix~\ref{sec:appendix_modelling}) show that such compact features can arise from scattering and viewing-geometry effects, though further modeling is needed to quantify the underlying conditions.

\subsection{No Jet Emission}
\label{sec:results_jet}

The combination of the WFC3/UVIS filters F606W and F814W has proven effective in revealing jets and distinguishing them from disk-reflected light in previous studies of EODs \citep[e.g.][]{Stapelfeldt+2014, Duchene+2024}. 
However, we do not find any evidence of jet emission in the HST observations of IRAS23077, which supports the conclusion by \citet{Monsch+2024} that the system is likely a more evolved Class II YSO with $\dot{M}_\mathrm{acc}\lesssim 10^{-8}\,\Msun/\mathrm{yr}$, rather than a younger, more strongly accreting Class~I or Class~F source. 

\section{Conclusions and summary}
\label{sec:conclusions}

We have presented new high-resolution (${\lesssim}0\farcs1$) HST/WFC3 observations of the recently identified, near edge-on protoplanetary disk IRAS23077 across six optical and near-IR broadband filters ($0.4$--$1.6\,\micron$).
These data offer the best spatially resolved view to date of the disk's vertical morphology and dust structure in scattered light. Our main findings are as follows:

\begin{itemize}

\item IRAS23077 shows the hallmark morphology of a highly inclined protoplanetary disk, with two bright lobes separated by a dark midplane lane. The disk appears flared and vertically extended, with faint scattered light reaching up to ${\sim}5''$ above the midplane. Prominent wispy structures of bright and dark material extend well above and in front of the disk, likely originating from its outer regions. Their irregular morphology suggests significant dynamical activity--possibly linked to turbulence, residual infall, or gravitational instability--indicating that the disk may not yet have reached vertical hydrostatic equilibrium.

\item A strong east–west brightness asymmetry is observed, with the western lobe being 6.5--$14.3\times$ brighter than the eastern one. This primarily reflects the disk’s ${\sim}80^\circ$ inclination, with additional wavelength dependence due to forward scattering and extinction.

\item The disk shows a clear wavelength-dependent vertical structure, with the dark lane and overall vertical extent decreasing toward longer wavelengths. This trend may result from opacity effects or true dust settling; however, comparison with radiative transfer models indicates that the available millimeter data lack the resolution needed to distinguish between these scenarios. A further limitation is the absence of resolved mid-IR observations, where intermediate-sized grains provide valuable constraints on vertical disk structure. Together, these factors prevent a definitive discrimination between the modeled scenarios. Future mid-IR imaging with JWST, and millimeter imaging with e.g., the ngVLA, would fill this wavelength gap and provide sufficient angular resolution toward improving vertical structure constraints for IRAS23077.

\item The HST imaging reveals filamentary structures extending up to ${\sim}10''$ above the northern side of the disk, with no southern counterparts. This intrinsic asymmetry suggests structural or illumination differences in the surrounding material.

\item Aperture photometry confirms a flux increase from optical to near-IR wavelengths, consistent with a PMS star surrounded by an IR-bright disk. NEOWISE monitoring reveals a mild ($\sim$0.3 mag) mid-IR brightening over the past decade and a weak ${\sim}0.1$\,mag W1-W2 color modulation. The origin of this variability remains uncertain but may be related to changes in disk illumination or variable obscuration.

\item A compact emission feature is detected near the disk center in the F160W image, coincident with the system's dynamical center. Its resolved size indicates that it is not direct starlight but scattered thermal emission from the inner disk, consistent with radiative transfer models.

\item No jet emission is detected in the HST data, implying that IRAS23077 is a more evolved Class~II, rather than a younger, strongly accreting Class~I source.
\end{itemize}

The HST observations presented here reveal that IRAS23077 is a structurally complex, dynamically active disk system, with wavelength-dependent vertical structure, filamentary asymmetries, and potentially ongoing or recent dynamical perturbations. However, critical gaps remain. In particular, no spatially resolved observations currently exist in the mid-IR---a key regime for probing intermediate-sized grains that are only partially coupled to the disk's gas. Without such data, it remains difficult to distinguish between competing models of dust growth and settling. Future JWST imaging and spectroscopy, combined with high-sensitivity (sub-)millimeter molecular line observations, will thus be essential to trace the kinematics of extended structures, test scenarios such as late infall or gravitational instability, and robustly characterize the vertical and radial dust distribution. Tailored hydrodynamical and radiative transfer modeling will be necessary to fully understand this exceptional edge-on disk system.

\begin{acknowledgments}
We thank the anonymous referee for a thorough and constructive report that improved the clarity and depth of this paper.
KM was supported by HST-GO-17751, JWST-GO-1905 and JWST-GO-3523.
JBL acknowledges the Smithsonian Institution for funding via a Submillimeter Array (SMA) Fellowship, and the North American ALMA Science Center (NAASC) for funding via an ALMA Ambassadorship.
Based on observations made with the NASA/ESA Hubble Space Telescope, obtained at the Space Telescope Science Institute, which is operated by the Association of Universities for Research in Astronomy, Inc., under NASA contract NAS5-26555. These observations are associated with program HST-GO-17751.
Support for program HST-GO-17751 was provided by NASA through a grant from the Space Telescope Science Institute, which is operated by the Association of Universities for Research in Astronomy, Inc., under NASA contract NAS5-26555.
Support for programs JWST-GO-1905 and JWST-GO-3523 was provided by NASA through a grant from the Space Telescope Science Institute, which is operated by the Association of Universities for Research in Astronomy, Inc., under NASA contract NAS 5-03127.
This research has made use of the NASA/IPAC Infrared Science Archive, which is funded by the National Aeronautics and Space Administration and operated by the California Institute of Technology.
This research made use of Photutils, an Astropy package for
detection and photometry of astronomical sources \citep{Bradley+2025_photutils}.
\end{acknowledgments}

%\appendix
%\input{05_appendices}


\vspace{5mm}
\facilities{HST (WFC3), MAST, IRSA, NEOWISE}


\software{\texttt{APLpy} \citep{RobitailleBressert2012}, \texttt{Astropy} \citep{astropy2013, astropy2018, astropy2022},  
\texttt{CMasher} \citep{cmasher},
\texttt{Matplotlib} \citep{Hunter2007}, 
\texttt{NumPy} \citep{Harris+2020}, 
\texttt{Photutils} \citep{Bradley+2025_photutils},
\texttt{SciPy} \citep{scipy},
\texttt{Seaborn} \citep{seaborn}
}

\appendix
\section{Radiative Transfer Modeling}
\label{sec:appendix_modelling}

\begin{table}[]
    \centering
    \caption{Model parameters adopted for our simulations of IRAS23077.}
    \begin{tabular}{c|cc}
        \hline
        \hline
        Parameter & Fixed/Varied & Value \\
        \hline
        $M_\star$ [$M_\odot$] & Fixed & 3.0 \\
        $R_\star$ [$R_\odot$] & Fixed & 4.9 \\
        $T_\star$ [K] & Fixed & 7400 \\
        \hline
        $\Sigma_c$ [g\,cm$^{-2}$] & Fixed & 0.02 \\        
        $\delta_{\rm cav}$ [] & Fixed & $10^{-6.5}$ \\        
        $R_{\rm c}$ [au] & Fixed & 600 \\        
        $R_{\rm cav}$ [au] & Fixed & 50 \\        
        $R_{\rm out}$ [au] & Fixed & 150 \\        
        $\gamma$ [] & Fixed & 1.0 \\        
        $\psi$ [] & Varied & 0.2, 0.5, 1.0, 1.5 \\        
        $a_{\rm small}$ [$\micron$] & Fixed & 0.1 \\
        $a_{\rm large}$ [$\micron$] & Fixed & 100 \\
        $\chi_{\rm S}$ [] & Fixed & 1.0 \\        
        $\chi_{\rm L}$ [] &Varied & 0.1, 1.0\\        
        $f$ [] &Varied & 0.80, 0.825, 0.85, 0.875, 0.90 \\
        \hline
        $d$ [pc] &Fixed & 300 \\
        $i$ [deg] & Varied & 77.5, 80., 82.5, 85. \\
        $PA$ [deg] & Fixed & 335 \\
        $\lambda$ [$\micron$] &Varied & 0.438, 0.606, 0.814, 1.05, 1.25, 1.60, 1300, 2700, 3100 \\        
    \end{tabular}
    \label{table:modelVals}
\end{table}

To test whether dust settling can explain IRAS23077's observed chromaticity, we performed radiative transfer simulations exploring its impact on the peak-to-peak intensity ratios and midplane thickness from optical to millimeter wavelengths. The models assume an axisymmetric disk and do not attempt to reproduce any of the observed disk substructures.

\begin{figure*}[t]
    \centering
    \includegraphics[width=1\linewidth]{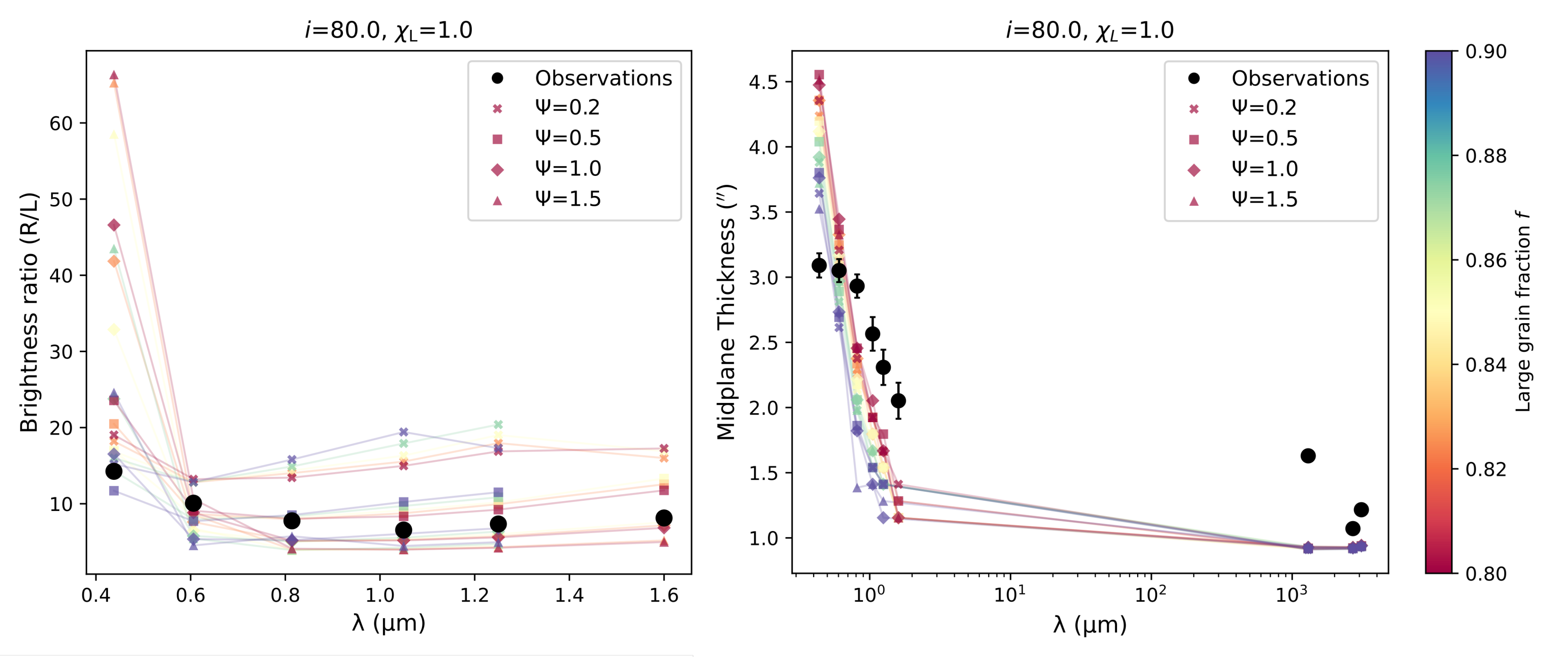}
    \caption{Comparison between simulated and observed disk properties. \textit{Left:} Peak-to-peak brightness ratios as a function of wavelength for the HST wavelengths. \textit{Right:} Midplane thickness as a function of wavelength, additionally including the millimeter data points. Black dots mark the observational measurements in both panels, while colored lines with different markers show the model realizations (flaring index $\psi$ and large grain fraction $f$) within the [$i=80^\circ$, $\chi_\mathrm{L}=1.0$] subset of our simulations. Although the simulations capture the overall trends, they do not reproduce the exact observed values, underscoring the limitations of our simplified model setup.}
    \label{fig:model_grids}
\end{figure*}

\subsection{Setup}
\label{sec:appendix_setup}

To model the dust emission of IRAS23077, we adopted a parametric protoplanetary disk model based on the transition disk prescription of \citet{Andrews+2011}. Although we did not attempt to reproduce the full SED of the disk and star \citep[as presented in][]{Berghea+2024}, we found that this transition disk framework provided a better description of the morphology and brightness of the long-wavelength emission. 

In this prescription, the disk is azimuthally and vertically axisymmetric, with a radial surface density profile that includes a dust-depleted cavity ($R_{\rm sub}<R\leq R_{\rm gap}$), a gap ($R_{\rm gap}<R\leq R_{\rm cav}$), and a denser outer disk ($R>R_{\rm cav}$, see Fig. 2 in \citealt{Andrews+2011} for a sketch). The surface density follows a viscous accretion disk profile with kinematic viscosity $\nu \propto R^\gamma$ \citep[where $\gamma=1$; cf.][]{Lynden-BellPringle1974, Hartmann+1998}, assuming the dust follows the behavior of the gas. The surface density distribution is given by:

\begin{equation}
    \Sigma_\mathrm{d} = \Sigma_{\rm c} \left( \frac{R}{R_\mathrm{c}} \right)^{-\gamma} \exp \left[ - \left( \frac{R}{R_\mathrm{c}} \right)^{2-\gamma} \right],
\end{equation}
with scaling radius $R_\mathrm{c}$ and normalization $\Sigma_{\rm c}=0.02\,\mathrm{g\,cm^{-2}}$. Inside the cavity region, an additional constant factor $\delta_{\rm cav}$ is multiplied by the surface density model. A full treatment of the dust-to-gas ratio is deferred to future work that will combine gas and dust observations of IRAS23077.

For the stellar properties, we adopted the spectral classification of \citet{Berghea+2024}, who identify the central star as type A9 based on optical scattered light spectroscopy. This spectral type corresponds to $T_{\rm eff}\approx 7400$\,K \citep[][Table~5]{PecaultMamajek2013}. From the \texttt{PARSEC} isochrones \citep{Bressan+2012} for a 2\,Myr PMS star at 300\,pc, this yields $L_\star\approx 65 L_\odot$ and $R_\star=4.9R_\odot$. For a dust sublimation temperature of 1500\,K \citep{Dullemond+2001}, we estimate $R_{\rm sub}\approx0.07(L_\star/L_\odot)^{0.5}$\,au, which corresponds to $R_{\rm sub}=0.56$\,au for IRAS23077.
We iteratively attempted to match the bulk radius of the disk in scattered light and millimeter data, as well as the millimeter brightness distribution, and found the combination of $R_{\rm gap}=50\,$au, $R_{\rm cav}=150\,$au, $R_{\rm c}=600$\,au and $\delta_{\rm c}=10^{-6.5}$ to provide a good approximation of the data. This was then used as our fiducial disk model.

The vertical disk structure is parameterized as a Gaussian distribution with a flaring law, $h = h_\mathrm{c}(R/R_\mathrm{c})^\psi$, where $h_\mathrm{c}$ is the dimensionless pressure scale height defined as $h_\mathrm{c} = c_\mathrm{s}/(R\Omega_\mathrm{k})$ for the midplane sound speed $c_\mathrm{s}$ and Keplerian frequency $\Omega_\mathrm{k}$, assuming a stellar mass of $3M_\odot$.

The dust is represented by two distinct populations: ``small" ($0.1\,\micron$) and ``large" grains ($100\,\micron$). The factor $f$ describes the fraction of large grains in the total dust density, leaving $(1-f)$ for the small grains. Vertical settling is parameterized by the disk scale height reduction factors $\chi_\mathrm{L}$ and $\chi_\mathrm{S}$ (where the subscripts ``S" and ``L" denote the small and large dust populations, respectively), which act as prefactors to the pressure scale height. In the unsettled case, $\chi_\mathrm{L}=\chi_\mathrm{S}=1$, while in a settled configuration large grains are reduced to $\chi_\mathrm{L}=0.1$ and small grains remain at $\chi_\mathrm{S}=1$. The vertical density profiles are then

\begin{equation}
l_{z,\mathrm{L}} = \frac{1}{\sqrt{2\pi}R\chi_\mathrm{L}h}\exp \left[ - \frac{1}{2}\left( \frac{z/R}{\chi_\mathrm{L}h} \right)^2 \right],
\end{equation}

\begin{equation}
l_{z,\mathrm{S}} = \frac{1}{\sqrt{2\pi}R\chi_\mathrm{S}h}\exp \left[ - \frac{1}{2}\left( \frac{z/R}{\chi_\mathrm{S}h} \right)^2 \right],
\end{equation}
with the 3D dust density distribution given by 
$\rho_\mathrm{L} = f\Sigma_\mathrm{d},l_{z,\mathrm{L}}$ for the large grains and $\rho_\mathrm{S} = (1-f)\Sigma_\mathrm{d}l_{z,\mathrm{S}}$ for the small grains.

We generated synthetic images of IRAS23077 using the Monte Carlo radiative transfer code RADMC-3D \citep{Dullemond2012_RADMC3D}, with dust density distributions as described above. Models were computed at the HST wavelengths 0.438, 0.606, 0.814, 1.05, 1.25, and 1.60\,$\micron$, and at SMA and NOEMA wavelengths of 1300, 2700, and 3100\,$\micron$. Three additional parameters are required for the synthetic observations: the disk position angle, set to $PA=335^\circ$ \citep{Lovell+2025}; the inclination $i$ (varied in our setup); and the system distance, assumed to be $d=300$\,pc. Dust grains were assumed to scatter isotropically and emit thermally. Table~\ref{table:modelVals} lists all parameters and their adopted values, indicating whether they were fixed or varied in our model. 

Since RADMC-3D produces full-resolution outputs, we convolved the synthetic images with the angular resolution of the observed data. For the HST observations, empirical PSFs were derived from field stars in the drizzled images, yielding FWHMs of $0\farcs09$, $0\farcs09$, $0\farcs10$, $0\farcs23$, $0\farcs24$, and $0\farcs25$ for F438W, F606W, F814W, F105W, F125W, and F160W, respectively. For the millimeter data, we adopted the restored clean beams: $0\farcs8\times0\farcs69$ ($BPA=61^\circ$) for SMA, and $0\farcs8\times0\farcs76$ ($BPA=50^\circ$) and $0\farcs81\times0\farcs77$ ($BPA=60^\circ$) for NOEMA at 112 and 96\,GHz, respectively \citep{Lovell+2025}.

Our investigation was designed specifically to test whether dust settling influences the observed disk morphologies. We therefore focused only on four parameters that most strongly affect the dust-lane thickness and scattered light lobe brightness ratios: the disk inclination $i$, the fraction of large grains composing the total dust population $f$, the settling factor of the large grains $\chi_\mathrm{L}$, and the vertical flaring index $\psi$. While this approach effectively probes the role of dust settling given the current data, the simplified model omits many features seen in the observations. A more comprehensive model of IRAS23077’s structure will therefore have to be developed in future work.

To probe the impact of grain settling, we considered two limiting cases: (i) a ``no-settling" model ($\chi_{\rm L}=1$), in which both grain populations are fully mixed vertically; and (ii) a ``10\% unsettled" model ($\chi_{\rm L}=0.1$), in which the large grains are confined to 10\% of the small-grain scale height, simulating their concentration toward the midplane. We further varied the fraction of large grains in the dust population $f$ between 0.80, 0.825, 0.85, 0.875, and 0.90.
For each of these setups we explored disk inclinations ranging from 77.5$^\circ$ to 82$^\circ$, and flaring indices of $\psi=0.2$, 0.5, 1.0, and 1.5, chosen to capture plausible degeneracies between brightness morphology and vertical extent. The broader range of $\psi=0.2$--1.5 compared to \citet{Andrews+2011} was motivated by recent scattered light studies of EODs \citep{Duchene+2024}. We find that increasing $\psi$ substantially enhances the extent of illuminated surface layers at large radii, strongly affecting the scattered light minor-axis profiles, while the corresponding impact on synthetic millimeter morphologies remains modest.

\begin{figure*}[t]
    \centering
    \includegraphics[width=1\linewidth]{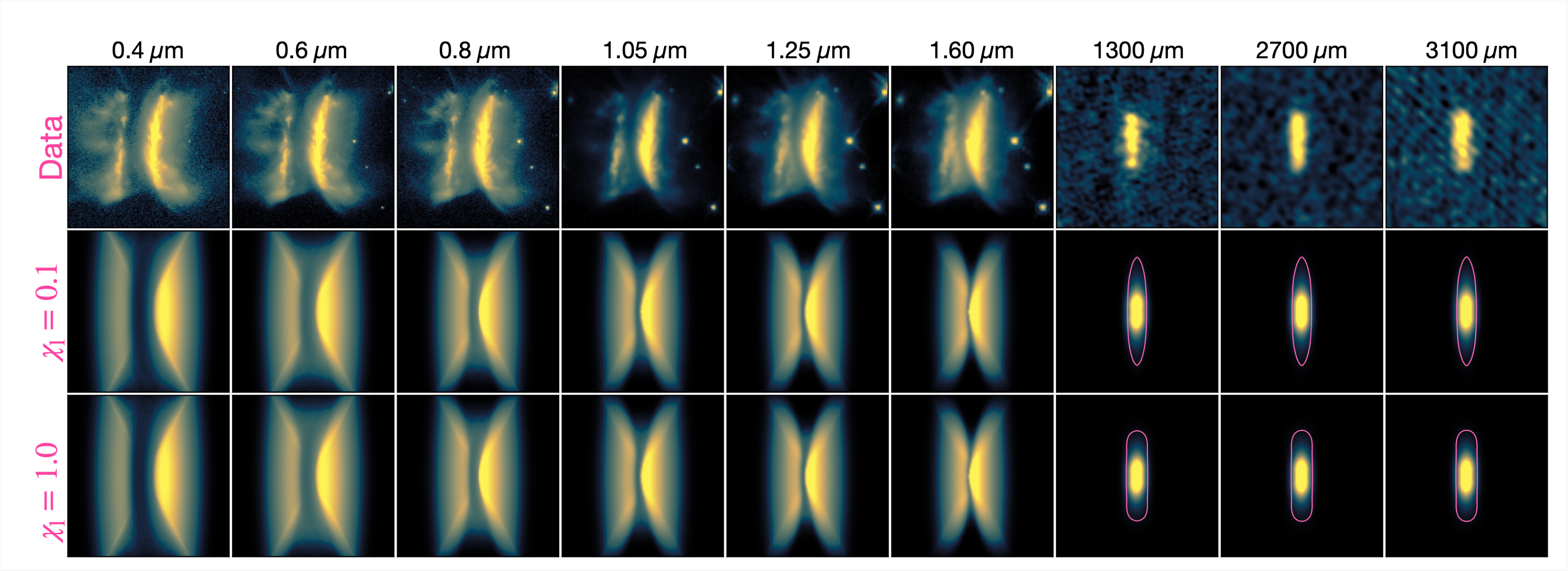}
    \caption{Comparison of observations and model results for one representative setup ($i=80^\circ$, $f=0.875$, $\psi=0.5$). The top row shows the HST, SMA, and NOEMA images ordered by wavelength. The middle row displays the results from the ``10\% unsettled" model ($\chi_\mathrm{L}=0.1$), and the bottom row those from the ``no-settling" model ($\chi_\mathrm{L}=1.0$). All simulations were convolved with the PSF corresponding to the observed image at the same wavelength (see \S\ref{sec:appendix_setup} for details). For the millimeter wavelengths, we also overplot contours at the 5\% peak-intensity level from the \textit{unconvolved} simulations, highlighting how the intrinsic emission morphology depends on the degree of dust settling.}
    \label{fig:obs_vs_model}
\end{figure*}

\subsection{Modeling Results}
\label{sec:appendix_results}

First, we tested whether our simulations are capable of reproducing the observed brightness ratios and midplane thicknesses of IRAS23077.
Figure~\ref{fig:model_grids} compares the observations (black dots) with a representative subset of our models ($i=80^\circ$, $\chi_\mathrm{L}=1.0$), where colored lines with different markers reflect the different model realizations (flaring index $\psi$ and large grain fraction $f$) within this subset. Although no single model simultaneously matches both observables, the simulations broadly reproduce the observed trend in brightness ratios, including their decline toward longer wavelengths. In contrast, the modeled midplane thicknesses are systematically smaller than observed, except at the shortest optical wavelengths ($0.4$–$0.6\,\micron$), where they are overestimated. 
These discrepancies likely reflect the simplifying assumptions of our modeling approach, such as axisymmetry, isotropic scattering, and an idealized dust structure, which limit the models' ability to reproduce the true midplane thicknesses of IRAS23077 across all wavelengths.
We also find significant parameter degeneracies, with comparably good fits to the observations arising from different combinations of model parameters. These residual discrepancies likely reflect the limitations of our simplified, axisymmetric modeling approach; reproducing the observations in full detail will, however, require more advanced, non-axisymmetric simulations.

Figure~\ref{fig:obs_vs_model} compares the observations to the results from one representative model setup ($i=80^\circ$, $f=0.875$, $\psi=0.5$) from the above subset. The second and third rows show the ``10\% unsettled” ($\chi_\mathrm{L}=0.1$) and ``no-settling" ($\chi_\mathrm{L}=1.0$) cases, respectively. As noted above, all simulations were convolved with the PSF corresponding to the observed image at the same wavelength.

The models broadly reproduce the overall morphology of IRAS23077 across all investigated wavelengths, although asymmetries such as shadowing and wisps were not modeled. Notably, the ``10\% unsettled" and ``no-settling" setups produce almost indistinguishable disk morphologies.
A similar outcome was reported by \citet{Duchene+2024} in their modeling of the EOD Tau042021: in the optical and near-IR scattered light, settled and unsettled models yielded very similar morphologies. At sub-mm wavelengths ($890\,\micron$/ALMA), however, they found a clear distinction---unsettled disks appeared significantly more flared, while settled disks produced narrowly confined, ``cigar-shaped" midplanes. In combination with high-resolution HST and JWST imaging, they concluded that in Tau042021 grains up to $\sim$10\,$\micron$ remain well coupled to the gas, whereas grains larger than $\sim$100\,$\micron$ are strongly settled toward the midplane.

In contrast, our convolved millimeter simulations of IRAS23077 do not show a flared morphology in either case, preventing us from discriminating between the two dust-settling scenarios. To illustrate that intrinsic differences do exist, however, we additionally overplot 5\% peak-intensity contours from the \textit{unconvolved} millimeter simulations, which reveal distinct morphologies for the settled and unsettled cases. However, these differences are washed out once the simulations are convolved with the beam, underscoring that at the resolution of our current data disk morphology alone cannot provide constraints on the degree of grain settling.

\section{NEOWISE Lightcurve}
\label{sec:appendix_neowise}

\begin{figure}
    \centering
    \includegraphics[width=0.8\linewidth]{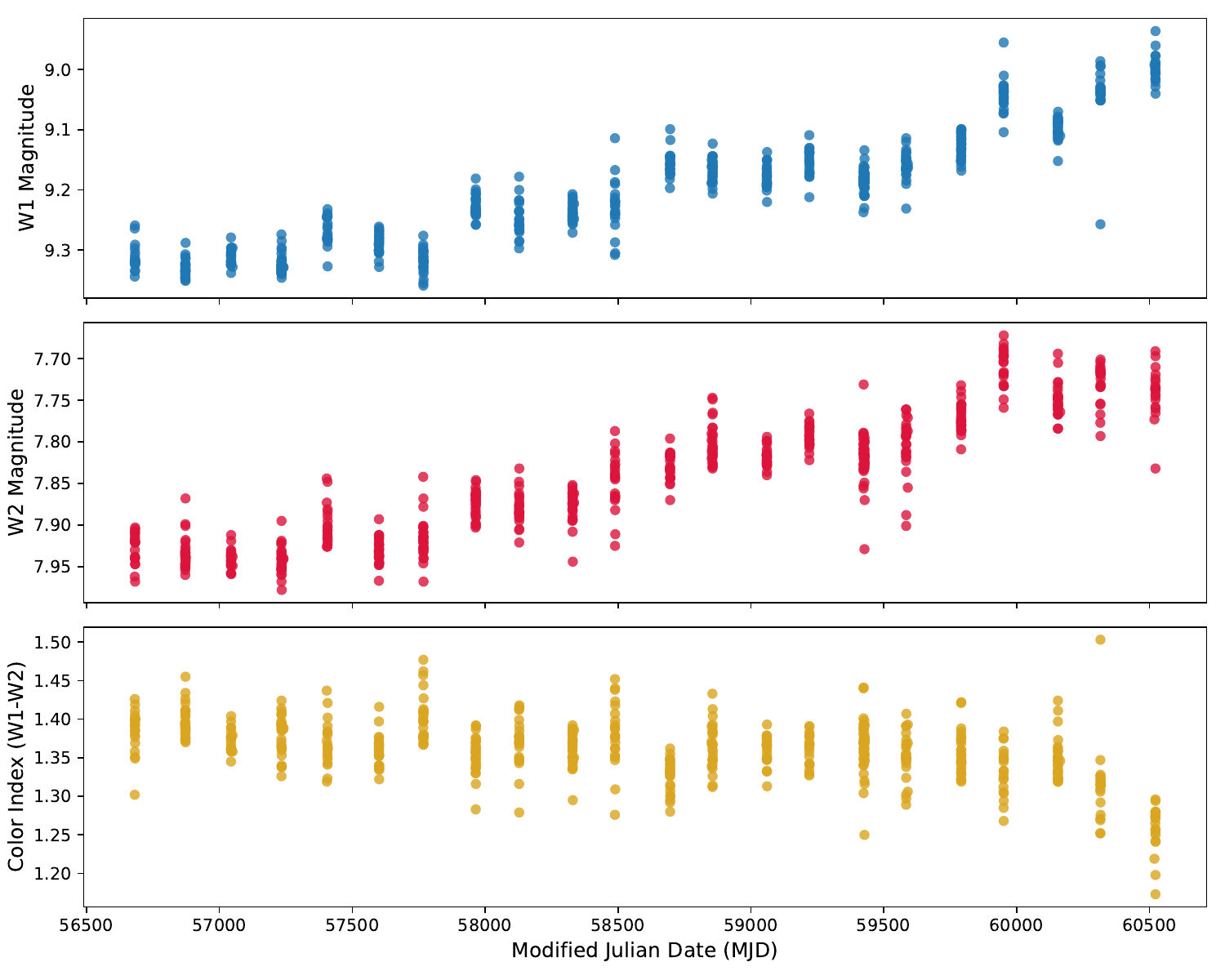}
    \caption{Lightcurve of IRAS23077 from the NEOWISE Reactivation Mission. \textit{Top:} W1 (3.4 $\mu m$) magnitude as a function of time. \textit{Center:} W2 magnitude (4.6 $\mu m$) magnitude. \textit{Bottom:} Color Index (W1-W2).}
    \label{fig:neowise}
\end{figure}

The Near-Earth Object Wide-field Infrared Survey Explorer \citep[NEOWISE;][]{Mainzer+2014} reactivated the WISE mission following the loss of cryogenic cooling, enabling nearly a decade of monitoring in the W1 and W2 bands. Although designed primarily for solar system science, these repeated observations allow us to assess possible near-infrared variability in IRAS23077. Figure~\ref{fig:neowise} shows the resulting light curves for $3.4\,\micron$ (W1) and $4.6\,\micron$ (W2) that are discussed in more detail in \S\,\ref{sec:results_photometry}.

\bibliography{literature}{}
\bibliographystyle{aasjournal}



\end{document}